\documentclass[conference]{IEEEtran}
\IEEEoverridecommandlockouts

\usepackage[utf8]{inputenc}
\usepackage[english]{babel}
\usepackage{graphics, graphicx}
\usepackage{amsmath, amsthm, amssymb, amsfonts}
\usepackage{algorithm}
\usepackage{algpseudocode}
\usepackage{multirow}
\usepackage{url}
\usepackage{soul}  
\usepackage{tabularx}
\usepackage{comment}
\usepackage{soul}
\usepackage{subfigure} 
\usepackage{tikz}
\usepackage{cite}
\usepackage{xcolor}
\usepackage{glossaries}
\usepackage{tabu}
\usepackage{tabularx}
\usepackage[nocomma]{optidef}
\usepackage{soul}

\usepackage{lipsum}  
\usepackage{xcolor, colortbl}
\usepackage{mathtools, nccmath}

\usepackage{xpatch}
\xpatchcmd{\NCC@ignorepar}{%
\abovedisplayskip\abovedisplayshortskip}
{%
\abovedisplayskip\abovedisplayshortskip%
\belowdisplayskip\belowdisplayshortskip}
{}{}

\newacronym{awgn}{AWGN}{additive white Gaussian noise}
\newacronym{ao}{AO}{alternating optimization}
\newacronym{bs}{BS}{base station}
\newacronym{bf}{BF}{beamforming}
\newacronym{chest}{CHEST}{channel estimation}
\newacronym{csi}{CSI}{channel state information}
\newacronym{dft}{DFT}{discrete Fourier transform}
\newacronym{dl}{DL}{downlink}
\newacronym{ee}{EE}{energy efficiency}
\newacronym{emf}{EMF}{electromagnetic field}
\newacronym[first= EMF exposure (EMFE)]{emfe}{EMFE}{electromagnetic field exposure}
\newacronym[first = Genetic Algorithm (GA)]{ga}{GA}{genetic algorithm}
\newacronym{jsac}{JSAC}{joint sensing and communication}
\newacronym{los}{LoS}{line-of-sight}
\newacronym{mimo}{MIMO}{multiple-input, multiple-output}
\newacronym{m-mimo}{M-MIMO}{Massive MIMO}
\newacronym{mmwave}{mmWave}{millimeter-wave}
\newacronym[plural = MPCs, firstplural = multipath components (MPCs)]{mpc}{MPC}{multipath component}
\newacronym{mrt}{MRT}{maximum ratio transmission}
\newacronym{mcs}{MCS}{Monte Carlo simulations}
\newacronym{nue}{NUE}{non-intended UE}
\newacronym{pa}{PA}{power allocation}
\newacronym{pl}{PL}{path-loss}
\newacronym[plural = RISs, firstplural = reflective intelligent surfaces]{ris}{RIS}{reflective intelligent surface}
\newacronym{ra}{RA}{resource allocation}
\newacronym{se}{SE}{spectral efficiency}
\newacronym{snr}{SNR}{signal-to-noise ratio}
\newacronym{tdd}{TDD}{time division duplex}
\newacronym[first = Uplink (UL)]{ul}{UL}{uplink}
\newacronym{ula}{ULA}{uniform linear array}
\newacronym{upa}{UPA}{uniform planar array}
\newacronym{ura}{URA}{uniform rectangular array}
\newacronym[plural = UEs, firstplural = users' equipment (UEs)]{ue}{UE}{user equipment}
\newacronym{5g}{5G}{5th-Generation}
\newacronym{6g}{6G}{6th-Generation}
\newacronym{3d}{3D}{three-dimensional}
\newacronym{2d}{2D}{two-dimensional}
\usepackage{geometry}
\definecolor{cian}{rgb}{.02,.7,.95}
\definecolor{gold}{rgb}{0.85,.66,0}
\definecolor{Gray}{gray}{0.95}

\newcommand{\bph}{\boldsymbol{\phi}}
\newcommand{\bPh}{\boldsymbol{\Phi}}

\newcommand{\brho}{\boldsymbol{\rho}}

\title{{EMF Exposure Mitigation in RIS-Assisted  Multi-Beam Communications}
\thanks{
This work was supported in part by the Coordenação de Aperfeiçoamento de Pessoal de Nível Superior - Brasil (CAPES) – Finance Code  001 and by the National Council for Scientific and Technological Development (CNPq) of Brazil under Grants 405301/2021-9, 141485/2020-5, and 310681/2019-7. R. Kotaba and P. Popovski were supported  in part by the H2020 RISE-6G project financed by the European Commission under grant no. 101017011. P. Popovski was supported by the Villum Investigator Grant “WATER” from the Velux Foundation. C. J. Vaca-Rubio was supported by the European Union's Horizon EUROPE research and innovation program under grant agreement No. 101037090 - project CENTRIC.
}
}
\author{
    \IEEEauthorblockN{
        {{Herman L. dos Santos}}\IEEEauthorrefmark{1},
        {{Cristian J. Vaca-Rubio}}\IEEEauthorrefmark{2},
        {{Rados\l{}aw Kotaba}} \IEEEauthorrefmark{2},
        {{Yi Song}}\IEEEauthorrefmark{3},
        \\
        {{Taufik Abrão}}\IEEEauthorrefmark{1},
        and {{Petar Popovski}}\IEEEauthorrefmark{2} \\
        }
        \IEEEauthorblockA{
        \IEEEauthorrefmark{1}\textit{\small Department of Electrical Engineering, Universidade Estadual de Londrina}, Londrina, Brazil\\
        \IEEEauthorrefmark{2}\textit{\small Department of Electronic Systems, Aalborg University}, Aalborg, Denmark\\
        \IEEEauthorrefmark{3}\textit{\small School of Electrical and Information Engineering, Zhengzhou University}, Zhengzhou, China \\
        \small E-mail: hermanlds@gmail.com, \{cjvr, rak, petarp\}@es.aau.dk, songyizzu@gs.zzu.edu.cn, and taufik@uel.br
        }
}

\begin{document}

\maketitle

\begin{abstract}
This paper proposes a method for reducing {third-party} exposure to electromagnetic fields (EMF) by exploiting the capability of a reconfigurable intelligent surfaces' (RIS) to manipulate the electromagnetic environment. We consider users capable of multi-beam communication, such that a user can use a set of different propagation paths enabled by the RIS. The optimization objective is to find propagation alternatives that allow to maintain the target quality of service while minimizing the level of EMF at surrounding non-intended users (NUEs). We provide an evolutionary heuristic solution based on Genetic Algorithm (GA) {for} power equalization {and} multi-beam selection of a codebook at the Base Station. Our results show valuable insights into how RIS-assisted multi-beam communications can mitigate EMF exposure with minimal degradation of the spectral efficiency.
\end{abstract}

\section{Introduction} \label{sec:intro}
{The rapid increase of the number of connected devices raises concerns about the \gls{emfe} from the signaling of devices under these technologies~\cite{Chiaraviglio2021}. As the \gls{6g} moves towards providing sustainable wireless connectivity,} it is imperative to explore innovative approaches to control EMFE~\cite{Airod2022}.

{Recent literature features several methods to deal with the \gls{emfe} concern {\cite{Phan2022, Ibraiwish2022,Emil2022,Zappone2022,HaoGuo2022}}.} {One promising direction for \gls{emfe}-aware communications in \gls{6g} systems relies on the concept of \gls{ris}, which consists of electronically controlled reflective elements that can passively perform signal phase-shifting. In this way, \gls{ris} acts as a system element that can manipulate the electromagnetic environment~\cite{Emil2022}.} Some studies include \gls{ris}-assisted \gls{ra} in \gls{emfe}-aware communications \cite{Ibraiwish2022,Zappone2022,HaoGuo2022}.

\gls{ul} \gls{emfe}-aware communications in \gls{ris}-assisted scenario have been studied in \cite{Ibraiwish2022}. The authors state that the \glspl{ue} are very exposed in \gls{ul} transmissions and propose a scenario exploiting the \gls{ris} reflective capabilities to increase the signal power received at the \gls{bs}. By jointly optimizing the combiner \gls{bf} vector, \gls{ris} coefficients, and transmit power, it is possible to significantly reduce the \gls{emfe}. Furthermore, the \gls{ris} can provide {energy-efficient} \gls{emfe}-aware communications \cite{Zappone2022} in \gls{ul} multi-antenna scenarios. It is shown that using the \gls{ris} it is possible to surpass the \gls{emfe} constraint and maintain fairly the energy efficiency.

{Besides self-\gls{emfe}-aware communications, in a real-world environment the \gls{ul} or \gls{dl} communication from a \gls{ue} might affect its surrounding in the sense of exposing other people to EMF radiation. We use the term \gls{nue} to denote those people, as they are potential users of the communication system at a future instant, but presently not communicating with the \gls{bs}.} The \gls{mmwave} have been employed in this scenario \cite{HaoGuo2022}. The orthogonality provided by the small wavelength turn possible transmitting in a two-path manner, \gls{bs}-\gls{ue} and \gls{bs}-\gls{ris}-\gls{ue}, and the authors show that it is possible to maximize the \gls{se} of a user when the direct path is constricted by the presence of a \gls{nue}. However, the method depends on high signaling {overhead}, \textit{e.g.} \gls{chest} to both \gls{ue} and \gls{nue}, and may be sensitive when the 
{low-exposure constraint }{appears} in both paths. Moreover, \gls{chest} routine can provide environment mapping \cite{YLin2022}, being able to localize the \glspl{nue} as scatterers, reducing the exposure and overhead of estimating the channel of these users.

\noindent\textit{Contributions}: {In this work, we address the problem of \gls{emfe}-aware communications by exploiting multi-beam communication to compose the \gls{bf} vector, thereby providing an effective solution for \gls{emfe} mitigation under requirements of \gls{se}. The solution is obtained via evolutionary heuristic techniques, namely \gls{ga}.} The results are evaluated 
for two scenarios: a) with perfect \gls{csi}; and b) scenario with only the localization information is available. 
The method is able to keep track of the \gls{emfe} under both assumptions.

\section{System Model} \label{sec:sysModel}

\begin{figure}[t]
\centering
\includegraphics[trim= 6mm 5mm 8mm 0mm,clip,width=.9\linewidth]{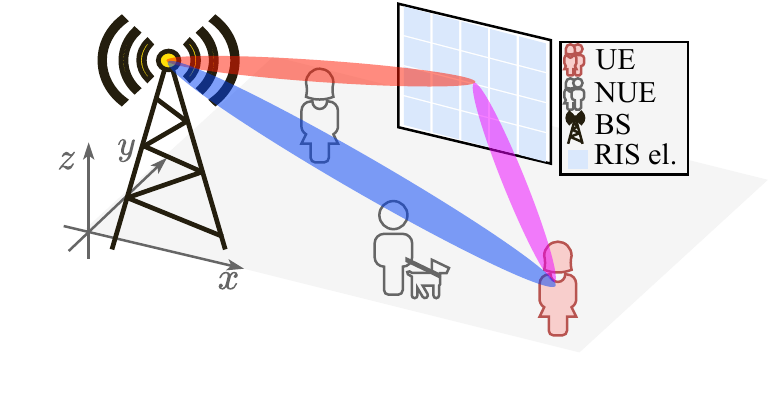}\vspace{-3mm}
\caption{\gls{mmwave} multi-beam \gls{emfe}-aware \gls{dl} system.}
\label{fig:system}
\end{figure}

Consider a \gls{mmwave} \gls{tdd} \gls{mimo} system where a \gls{bs} equipped with $M$ antennas communicates in \gls{dl} with a single-antenna \gls{ue} aided by an \gls{ris} composed of $N$ reflective elements. Both \gls{bs} antennas and \gls{ris} elements are arranged in a rectangular \gls{ula} fashion with $d=\lambda/2$ meters spacing, where $\lambda$ is the carrier wavelength. The elements are positioned such that the normal of \gls{bs} antennas is aligned with the $x$-axis, \textit{i.e.} $M_y=M_z$, and \gls{ris} elements are distributed with $N_x=N_z$. The \gls{bs} communicates with the \gls{ue} using a \gls{bf} vector:
\useshortskip
\begin{equation}
\mathbf{w} = \sum_{m=1}^M \sqrt{\rho_m} {\mathbf{w}_m} \in \mathcal{C}^{M \times 1}
\end{equation}
where {$\mathbf{W}=[{\bf w}_1, {\bf w}_2, \ldots, {\bf w}_M] \in \mathcal{C}^{M \times M}$} is the codebook and $\brho$ the \gls{pa} vector. A representation of the system is depicted in Fig. \ref{fig:system}. In such a case, the \gls{bs} can choose to exploit the direct, reflected paths, or multiple beams to compose the signal. The \gls{dl} transmitted symbol $\mathbf{y}$ is represented as 
\begin{equation}\label{eq:receivedSignal}
    \mathbf{y} = \mathbf{w}^H x = 
\left(\sum\limits_{m=1}^M\sqrt{\rho_m} \mathbf{w}_m^H\right) x  \, {\in \,\, \mathcal{C}^{M \times 1}}
    \end{equation}
\noindent 
where $x$ is the transmitted complex symbol, which has unitary power, \textit{i.e.}, $\mathbb{E}[|x|^2]=1$ and $[\cdot]^H$ denotes the Hermitian transpose operator. 
The {\it reciprocal channel} of the $k$-th \gls{ue} is composed by the reflected and direct paths:
\begin{equation}
\label{eq:chSmallScale}
\mathbf{h}_k = \underbrace{{\mathbf{H}_B^{R}}^H\bPh\mathbf{h}_R^{k}}_{{\rm reflected \ path}} + \underbrace{\mathbf{h}_B^k}_{\rm direct \ path},
\end{equation}
\noindent where $\mathbf{H}_B^{R} \in \mathbb{C}^{N\times M}$ is the \gls{bs}-\gls{ris} channel, $\mathbf{h}_R^{k} \in \mathbb{C}^{N\times 1}$ and $\mathbf{h}_B^{k} \in \mathbb{C}^{M\times 1}$ are the \gls{ris}-\gls{ue} and \gls{bs}-\gls{ue} channel vectors, respectively. {The matrix} $\bPh = {\rm diag} (\bph) \in \mathbb{C}^{N\times N}$ is a diagonal matrix of \gls{ris} elements phase-shift, and $\phi_{i,i}=e^{j\phi_i} \in [-\pi;\pi]$ denotes the $i$-th diagonal entry. Along with the active \gls{ue}, there is a set of \glspl{nue} that are not communicating with the \gls{bs} and {\it should not be exposed to excessive} \gls{emf}. {The \gls{emfe} is defined as the power density that impinges in a specific area, which is discussed further in this paper.} 
The channel \gls{bs}-\glspl{nue} follows the same rule as applied to \gls{ue}.

We adopt the Rayleigh channel model \cite{Goldsmith2005} with the channels coefficients described as a factor of the \gls{pl}, specifically mmMAGIC urban \gls{los} model \cite{Rappaport2017}, Eq. \eqref{eq:mmMAGICPL}, and small-scale fading configuring multi-path propagation. The \gls{pl} is dependent of \gls{3d} Euclidean distance $d_{3D}$ in meters, the carrier frequency $f_c$ in GHz, and stochastic shadow fading {standard deviation} $\sigma_{\rm SF}$ {in dB:}
\begin{equation}\label{eq:mmMAGICPL}
    \beta = 19.2 \log_{10}(d_{3D}) + 32.9 + 20.8\log_{10} (f_c) + \sigma_{\rm SF}.
\end{equation}
This \gls{pl} model is valid for the range of $6$ to $100$ GHz and assumes \gls{los} propagation. Hence, the channel is a product of the \gls{pl} with random amplitudes and phases normally distributed, unitary variance, from small-scale fading {modeling}. We evaluate the signal power impinging in an arbitrary point in space $\mathbf{q} \in \mathbb{R}^{3 \times 1}$ as\footnote{A point $\mathbf{q}$ may represent a \gls{ue}, \gls{nue}, or compose a set denoting a group of \glspl{nue}.}
\begin{equation}
\label{eq:pointchannel}
\begin{split}
|\mathbf{y}_q|^2 = & |\mathbf{w}^H({\mathbf{H}_B^R}^H \bPh \mathbf{h}_R^q + {\mathbf{h}^q_B})|^2.
\end{split}
\end{equation}

\section{EMFE-aware communications}
{The objective of our design is twofold.} We would like to increase, or at least maintain the quality of service for the \gls{ue}, while limiting the \gls{emfe} at points/areas where the \glspl{nue} are located. This is achieved by considering multiple propagation paths (besides the direct and reflected paths), which offers a bigger solution space. For the sake of tractability, we focus on the designs exploiting mutually orthogonal beams, \textit{i.e.} not interfering with each other, which further allows us to apply power equalization along with beam selection. Throughout, we assume the \glspl{nue}' positions are known.\footnote{Localization methods are out of scope herein. Future works will encompass methods and performance under localization errors.}


\subsection{Codebook design and Spectral Efficiency }

The ability to manipulate the electromagnetic environment with \gls{ris} adds another degree of freedom to communication systems. Because the reflected path can be controlled, it becomes a viable solution for \gls{emfe} constrained communications \cite{HaoGuo2022}, in addition to the direct path between \gls{bs} and \gls{ue}.
To that end, in this work we examine both {\it direct and reflected paths} while taking into account a {\it threshold for the power allocated to the direct one}. This threshold is determined using \gls{mrt} and is necessary due to the presence of the \gls{nue} between the \gls{bs} and \gls{ue}. The remaining power is then allocated to the reflected path.
However, such an approach is very sensitive to the quality of the \gls{chest} routine. To overcome this issue, we choose a \gls{dft} codebook-based approach, defining a {\it set of orthogonal beams} obtained by the \gls{dft} used to communicate with the \gls{ue}. 
The codebook of orthogonal codewords used by both \gls{bs} and \gls{ris} is defined as:
%
\begin{equation}
    \mathbf{W}^I = \{\boldsymbol{\mu}_i\}^{\sqrt{I}}_{i=1} \otimes \{\boldsymbol{\nu}_j\}^{\sqrt{I}}_{j=1},
\end{equation}
\noindent with $I\in \{N,M\}$ and
\begin{align}
    \boldsymbol{\mu}_i = \left[1, e^{\frac{2\pi(i-1)}{\sqrt{I}}}, \dots , e^{\frac{2\pi(i-1)(\sqrt{I}-1)}{\sqrt{I}}}\right],\\
    \boldsymbol{\nu}_j = \left[1, e^{\frac{2\pi(j-1)}{\sqrt{I}}}, \dots , e^{\frac{2\pi(j-1)(\sqrt{I}-1)}{\sqrt{I}}}\right].
\end{align}
where the operator $\otimes$ is the  
Kronecker product. 

Using Shannon's capacity equation \cite{Marzetta2016}, 
the \gls{ue} \gls{se} is obtained by  considering the ratio of impinging signal power, the receiver noise power $\sigma^2_w$, and the 
bandwidth $B$, also considering 
\gls{bf} vectors $\boldsymbol{\rho}$ as:
\begin{align}\label{eq:SE}
    {\rm SE} = B\mathrm{log}_2\left(1 + \frac{|\mathbf{w}^H({\mathbf{H}_B^R}^H \bPh \mathbf{h}_R^k + {\mathbf{h}^k_B})|^2}{\sigma^2_w}\right),
\end{align}
\noindent where $\sigma_w^2 = B\cdot N_0\cdot N_f$, being $N_0$ the noise power {spectral density} and $N_f$ the noise figure. 

\vspace{2mm} 
\noindent{\bf EMFE-aware optimization}: The \gls{bs} equalizes the power between the beams available in $\mathbf{W}$ assigning values to $\brho$. The \gls{emf} manipulation is proportional to the power of the signal that impinges on the target points in set $\mathcal{Q} = \{q_1, q_2, \dots, q_{|\mathcal{Q}|}\}$, being ${|\mathcal{Q}|}$ the cardinality of the set. The problem is subject to a minimum \gls{se} to the \gls{ue}. Hence, the power density $\bar{P}$ {in [W/m$^2$] or [W/Area-$\mathcal{Q}$] over a specific area such that} the sum of the power density in each point $q \in \mathcal{Q}$ {can be expressed as}: 
%
\begin{align}\label{eq:pointpower}
 \bar{P} = \frac{1}{|\mathcal{Q}|}\sum\limits_{q\in\mathcal{Q}} \left|\mathbf{w}^H ({\mathbf{H}_B^R}^H \bPh \mathbf{h}_R^q + \mathbf{h}^q_B) \right|^2.
\end{align}

\noindent We can formally describe the \gls{emfe}-aware communications as a minimization of power density $\bar{P}$ over a set of points $\mathcal{Q}$, Eq. \eqref{eq:pointpower}, under the \gls{emfe} and physical constraints of the system. This problem is formulated as\useshortskip
\begin{subequations}
    \begin{align}
    {(\mathcal{P}1)}  \quad \underset{\brho,\bph }{\mathrm{minimize}} \quad& \bar{P} \\
    \mathrm{s.t.} \quad& \mathrm{SE} \ge \mathrm{SE_{min}},\label{eq:ueQoS}\\ 
     & \mathbf{w} \in \mathbf{W}^B \label{eq:bscbCons}\\
     & \bph \in \mathbf{W}^R \label{eq:riscbCons}\\
     &  \sum\limits_{i=1}^M \rho_i \leq P \label{eq:powerCons} 
    \end{align}
\end{subequations}
wherein \eqref{eq:ueQoS} expresses the constraint of guaranteed
\gls{se} to the \gls{ue}, \eqref{eq:bscbCons} limits the \gls{bs} \gls{bf} vectors to concatenating the codebook words, while \eqref{eq:riscbCons} implies the usage of a vector in \gls{ris} codebook as reflection coefficients. Finally, \eqref{eq:powerCons} formally delimits the power pool constraint.  
{It is worth noting that {($\mathcal{P}1$)} is challenging to solve due to the non-convexity of the objective function and constraints.} 

\subsection{\gls{chest} and Localization Impairments}\label{sec:chestvsloc}
The problem $(\mathcal{P}1)$ can achieve global minima if the \gls{csi} have been perfectly estimated, which is a weak assumption since the \gls{chest} process can be quite costly in terms of resources and signaling. 
An alternative solution, requiring less overhead, is the estimation of the positions of the environmental elements through sensing.

For \gls{chest} routine, $\kappa M$ and $\kappa MN$ channel coefficients need to be estimated for direct and reflected paths, respectively, where $\kappa \triangleq K + \tilde{K}$ denotes $K$ \glspl{ue} and $\tilde{K}$ \glspl{nue}. Based on the three-phase \gls{chest} framework \cite{Zhaorui2020}, the minimum signaling consisting of
$$
\vspace{-2.5mm}
\kappa+N+\max\left\{\kappa-1,\,\, \lceil(\kappa-1)\cdot N/M\rceil\right\}\,\, \,\, [\text{pilots}]
$$ 
is necessary just for perfectly recovering \gls{csi}, 
and that is under the assumption of no receiver noise, which is unrealistic. Hence, perfect \gls{csi} is an idealized case but is employed to derive upper bounds of system performance.



In contrast to traditional \gls{ris} \gls{chest} procedures that require the cascaded channel estimation at either the \gls{bs} or the \gls{ue}, several recent works use the localization of the \glspl{ue} \cite{jiang2022optimization, keykhosravi2022leveraging} in \gls{jsac}. \gls{jsac} allows retrieving environmental information along with the \gls{nue} position, which suits the case of \gls{emfe} mitigation since the same signaling used to localize the \gls{ue} of interest can be employed to retrieve the \glspl{nue} position. For instance, the proposed framework in \cite{jiang2022optimization} reduces pilot overhead and the need for frequent RIS reconfiguration by optimizing the configuration of the RIS for several channel coherence intervals. This leads to a novel frame structure, comprising infrequent localization and RIS control tasks, combined with a more frequent optimization of the BS precoders to maximize the transmission rate. This approach leverages accurate location information obtained with the aid of several RISs as well as novel RIS optimization and 
CHEST methods. These works further motivate our method, in which by obtaining the users' location, we can avoid the dense overhead related to the \gls{chest} {procedure}.

\section{Proposed Solutions}
\label{sec:Solutions}
We study and compare two approaches to solve the {EMFE optimization} problem; \textit{a)} the rate maximization constrained by \gls{emfe} framework originally presented in \cite{HaoGuo2022}, namely \gls{dft}-based beam and power optimization; \textit{b)} employment of evolutionary heuristic, specifically \gls{ga}, to exploit the solution space of combining more than two beams.

\vspace{2mm}
\noindent\textbf{\gls{mrt} and \gls{dft} beam (MRT-DFT) \cite{HaoGuo2022}}: assuming perfect \gls{csi} of both \gls{ue} and \gls{nue}, obtaining a quasi-optimal and low complexity solution. The method is set as follows. The direct beam, $\mathbf{w}_d$, follows a \gls{mrt} rule, taking its' coefficients as the inverse of the direct \gls{bs}-\gls{ue} channel, and {another jointly optimizes the \gls{bf} vector towards the \gls{ris} and its' reflection coefficients by extracting the word with the maximum gain and matching the beam $\mathbf{w}_r$. Hence, the power of each beam is weighted to maximize the \gls{se} under a constraint of maximum impinging power at the \gls{nue}. We describe this implementation in Algorithm \ref{ag:baseline}.}

\begin{algorithm}
\caption{\gls{mrt} and \gls{dft} beam\cite{HaoGuo2022}} 
\begin{algorithmic}[1]
\Require{$\mathbf{W}^R,\mathbf{H}_B^R,\mathbf{h}_B^{k},\mathbf{h}_R^{k},\mathbf{h}_B^{{q}},\mathbf{h}_R^{{q}},P,\bar{P}$}
\Ensure{$\mathbf{w}_d,\mathbf{w}_r,\bPh,{\rho}_d,{\rho}_r$}
    \State Define $\mathbf{w}_d \leftarrow  {\mathbf{h}_B^k}^H / \|\mathbf{h}_B^k \|$
    \For{$i=1$ to $N$}
             \State $\bPh \leftarrow  {\rm diag}(\mathbf{W}^i_{R})$ 
             \State Evaluate $|{\mathbf{H}^R_B}^H\bPh \mathbf{h}^k_R |^2$
             \State Store the reflected path power gain under $\mathbf{W}_R^i$
    \EndFor
    \State Assign the beam with the highest gain to $\bPh$;
    \State Define $\mathbf{w}^H_r \leftarrow  ({\mathbf{H}^R_B}^H\bPh \mathbf{h}^k_R )^H/\|{\mathbf{H}^R_B}^H\bPh \mathbf{h}^k_R \|$
    \State $\rho_d \leftarrow  {\bar{P}} / {\|\mathbf{w}^H_d \mathbf{h}^{q}_B\|} $
    \State $\rho_r \leftarrow  {P} - \rho_d$
    \end{algorithmic}
    \label{ag:baseline}
\end{algorithm}

The implementation presented in Algorithm \ref{ag:baseline} assumes perfect \gls{csi} for both \gls{ue} and \gls{nue}. The performance and reliability are equivalent to the \gls{chest} results. It is pointed out that another drawback is when both direct and reflected path is virtually obstructed let's say by \glspl{nue}, limiting the amount of power that can be employed. We choose to exploit such characteristics in our numerical results.

\vspace{2mm}

\noindent\textbf{Multi-Beam Power Equalization by \gls{ga} (MB-GA)}: we propose the use of a \gls{ga} in $M$-beam search-and-weighting defined in codebook $\mathbf{W}^B$. We choose this method because the non-linear, non-convex, mixed integer-quadratic programming problem of $\mathcal{P}1$ is not solvable {in a polynomial time}, 
{while} \gls{ga} has a widespread in the search space. Generally, this method {attains} at least quasi-optimal solutions when the problem is well-defined by {a cost function.} The authors highlight that there are several methods for solving ($\mathcal{P}1$), and will be investigated in future works.
{The \gls{ga} procedure} is performed as \cite{PegoraraSouto2019}:

\noindent{\bf 1.} \textit{Creating Initial Population:} a number of initial population \texttt{nP} of chromosomes $\mathcal{X}^0 =\{\mathbf{x}_1, \mathbf{x}_2, \dots, \mathbf{x}_{\texttt{nP}}\}$, \textit{i.e.} individual candidate solutions, Eq. \eqref{eq:GAinitPop}, is uniformly generated in feasible solution space. We generate a batch of solutions under the constraints defined in equations (11d) and (11e), check their feasibility by evaluating Eq. (11b), and re-generate those that do not attend to the constraints. These entries are organized as \useshortskip
\begin{equation}\label{eq:GAinitPop}
    \mathbf{x}_i = [\rho_1, \rho_2, \dots, \rho_M, \boldsymbol{\omega}^R]^T, \quad  i = \{1, 2, \dots, \mathtt{nP}\}, \vspace{-1mm}
\end{equation}
\noindent being $\boldsymbol{\omega}^R \in \{0,1\}^{N\times 1}, \sum \boldsymbol{\omega}^R = 1$, and $\bPh = \mathbf{W}^R \boldsymbol{\omega}^R$.

\noindent{\bf 2.} \textit{Scoring and Scaling of Population:} the solutions are scored as their problem-specific fitness function, Eq. \eqref{eq:pointpower}, value\footnote{As the \gls{ga} is applied {in the cost function} minimization, the lower the value of the fitness function, the better is the candidate-solution.} $\mathcal{Y}^j = f(\mathcal{X}^j)$, being $j$ the generation index. Then, the solutions are escalated to ensure those with higher quality are most likely chosen as the next generation of parents. Formally, the escalated solution in $j$-th iteration is defined as \useshortskip
\begin{equation}\label{eq:GAescEq}
     \bar{\mathcal{Y}}^j= g(\mathcal{Y}^j) = \alpha \mathcal{Y}^j
\vspace{-3mm}
\end{equation}
\noindent being $\alpha$ the scaling function, which is a design parameter.

\noindent{\bf 3.} \textit{Elite Retaining and Tournament:} the elite count \texttt{eC} parameter defines how many individuals are guaranteed to advance to the next generation after scaling. The \texttt{eC} solutions with the lowest fitness function values are parents for the offspring. The rest of the parents are selected by a {\it tournament function}, which randomly selects \texttt{nT} individuals, defining which \textit{survives} {by applying} a probabilistic test. 
Formally, the probability of the $k$-th individual proceeding in the tournament is: \useshortskip
\begin{equation}\label{eq:GATourn}
    p_k=\frac{\bar{y}_k}{\sum_{i=1}^{{\mathtt{nT}}}\bar{y}_i}, \qquad k \in \mathtt{nT}
\vspace{-2mm}
\end{equation}
\noindent where 
{$\mathtt{nT}$} is a parameter that defines how many individuals besides the elite are allowed in the next generation. The probability of an individual surviving a generation is proportional to the quality of its solution.

\noindent{\bf 4.} \textit{Crossover and Mutation:} 
At the end of a generation, the crossover function is applied  in the set of parents, including the elite and tournament winners, and the mutation function in the offspring generation. The \texttt{nT} individuals left from elite count plus tournament result are input to the crossover function, which creates the children by combining the values of two solutions, let's say $\mathbf{x}_i$ and $\mathbf{x}_j$, $\bar{y}_i < \bar{y}_j$, and multiplying by a random value. Formally, the child $\bar{\mathbf{x}}$ which belongs to the next generation is defined as \useshortskip
\begin{equation}
\bar{\mathbf{x}} = \mathbf{x}_i + \mathcal{U}[0,1] (\mathbf{x}_i - \mathbf{x}_j). \vspace{-2mm}
\end{equation}
\noindent After defining the crossover, the {\it mutation} process is applied. The mutation occurs under a probability $p_m$, also a parameter of design, and replaces an entry of the child solution with a random value. The mutation defines the end of a generation, then the algorithm goes back to Step 2 and proceeds until a stopping {criterion is satisfied}. 

\section{Simulation Results}
We carry out \gls{mcs} to evaluate and compare with the reference method \cite{HaoGuo2022}, which assumes perfect knowledge of the \gls{csi} and does not exploit the sensing capabilities of \gls{ris} for location estimation. We evaluate the methods with one and two \glspl{nue} in four scenarios, as depicted in Figure \ref{fig:system}, where the NUE$_1$ represents the one in \gls{bs}-\gls{ue} line (person with a dog, in the graphical representation), and NUE$_2$ in \gls{bs}-\gls{ris} line. 
{Four configurations are evaluated}:
\begin{itemize}
\item Scenario 1.1: There is one \gls{nue} moving along the line between the \gls{bs} and \gls{ue}. We assume the \gls{bs}, \gls{ris},  \gls{nue} and \gls{ue} are all at the same height $z$.
\item Scenario 1.2: There is one \gls{nue} moving along the line between the \gls{bs} and \gls{ue}. We assume the \gls{bs}, \gls{ris},  \gls{nue} and \gls{ue} where the different heights follow the rule \gls{bs}$_z>$ \gls{ris}$_z>$ \gls{ue}$_z=$ \gls{nue}$_z$.
\item Scenario 2.1:  There are two \glspl{nue} moving along the line between the \gls{bs} and \gls{ue}, and the line between the \gls{bs} and \gls{ris}. We assume the \gls{bs}, \gls{ris},  \gls{nue} and \gls{ue} are all at the same height $z$.
\item Scenario 2.2: There are two \glspl{nue} moving along the line between the \gls{bs} and \gls{ue}, and the line between the \gls{bs} and \gls{ris}. We assume the \gls{bs}, \gls{ris},  \gls{nue} and \gls{ue} {with} different heights {following} the rule \gls{bs}$_z>$ \gls{ris}$_z>$ \gls{ue}$_z=$ \gls{nue}$_z$.
\end{itemize}
Scenarios 2.1/2.2 are more challenging due to the limitation in the amount of power that can be allocated to the direct path. To ensure a fair comparison, we set \gls{se}$_{\min}$ equals to the \gls{se} obtained in \gls{dft}-based \gls{bf} scheme. For both methods, we consider the selected beams leak some energy, which increases the evaluated exposure and \gls{se}.
We will evaluate the results under two use cases \textit{i)} the ideal assumption of perfect knowledge of the \gls{csi}, and \textit{ii)} the estimation via the \glspl{ue} and \glspl{nue} locations. We detail the parameters of our simulation in Table \ref{tab:SimParams}.

\begin{table}[htbp]
\caption{Simulation Parameters}
\label{tab:SimParams}
\centering
\begin{tabular}{llr} \hline
\rowcolor{Gray}\multicolumn{3}{c}{System}                     \\ \hline
\textbf{Parameter }       & \textbf{Value}      &             \\
BS Ant.                   & $36$                &             \\
RIS El.                   & $100$               &             \\
\multirow{2}{*}{BS Pos.}  & I.  $[-80, 0, 1.5]$ & [m]         \\ 
                          & II. $[-80, 0, 10]$  & [m]         \\
\multirow{2}{*}{RIS Pos.} & I   $[0, 50, 1.5]$ & [m]         \\
                          & II. $[0, 50, 5]$   & [m]         \\
UE Pos.                   & $[80, 0, 1.5]$      & [m]         \\
Carrier freq.             & $f_c = 28 $         & [GHz]       \\
Noise power               & $N_0 = -174$        & [dBm/Hz]    \\ 
Noise figure              & $N_f = 10$          & [dB]        \\
Bandwidth                 & $B = 100$           & [MHz]       \\
Available power           & $P = 43$            & [dBm]       \\ 
Shadow Fading             & $\sigma_{\rm SF} = 2$& [dB]       \\ \hline
\rowcolor{Gray}\multicolumn{3}{c}{Genetic Algorithm}          \\ \hline
Population size           & $\mathtt{nP} = 300$ &             \\
Elite counter             & $\mathtt{eC} = 2$   &             \\
Tournament size           & $\mathtt{nT} = 4$   \\
Scaling function          & $\alpha$ rank based \\
Mutation probability      & $p_m = 0.05$        &           \\
Constraints tolerance     & $10^{-6}$           &             \\ \hline
Realizations              & 1000                & MCS\\  
\hline
\end{tabular}
\end{table}


\vspace{-2mm}
\subsection{Performance under Perfect \gls{csi}}
Here we evaluate the performance of the method with perfect \gls{csi} information for the described scenarios.

\begin{figure}[bp]
\centering
\includegraphics[trim= 0mm 0mm 0mm 0mm, clip, width = .95\linewidth]{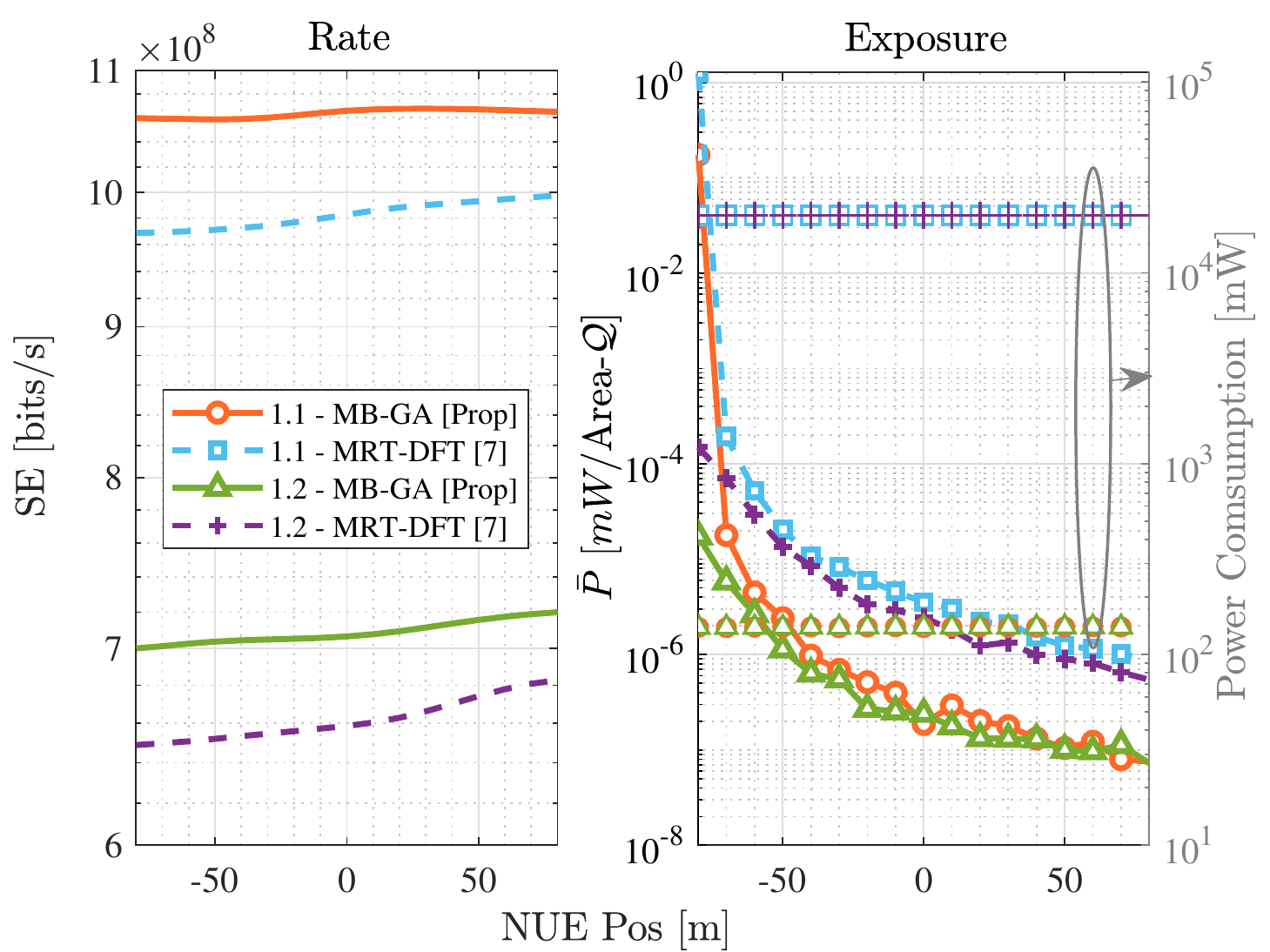}
\caption{Achieved rate \textit{(left)} and exposure \textit{(right)} performance under perfect \gls{csi} for Scenario 1.1 and 1.2. The \gls{nue} is positioned in the origin of the $y$-axis and moves from $-80$ meters (\gls{bs} position) to $80$ meters (\gls{ue} position). }
\label{fig:Scen1PCSI}
\end{figure}

\noindent\textbf{Scenarios 1.1 and 1.2}: Scenario 1 represents straightforward the baseline scenario. There is only \gls{nue}$_1$ performing \textit{virtual} blockage in the direct path. Intuitively, this scenario is the one that brings the performance upper bound in terms of achievable \gls{se} since the reflected path can be fully exploited. The performance under perfect \gls{csi} for both methods is presented in Fig. \ref{fig:Scen1PCSI}.

Fig. \ref{fig:Scen1PCSI} shows that the larger search space provided in our proposal (solid lines) is capable of improving the \gls{se} while achieving low exposure communication
. For both methods, we evaluated the performance considering the instantaneous coefficients of the channels were known to the \gls{bs}. Exploiting the higher search space reveals that our method can compose a multi-beam transmission coherently sums to the \gls{ue} channel, surpassing the \gls{mrt}{-\gls{dft} method}. 
Besides improving the rate of the \gls{ue}, the composition of beams also can mitigate more efficiently the {EMF}, 
being more effective. The reference method utilizes all the power pool, {\it i.e.}, $\rho_d + \rho_r = P$, while the proposed method uses {$0.79$\%} of the available power, presenting an improvement in power allocation. 

\begin{figure}[htbp]
\centering
\includegraphics[width = .95\linewidth]{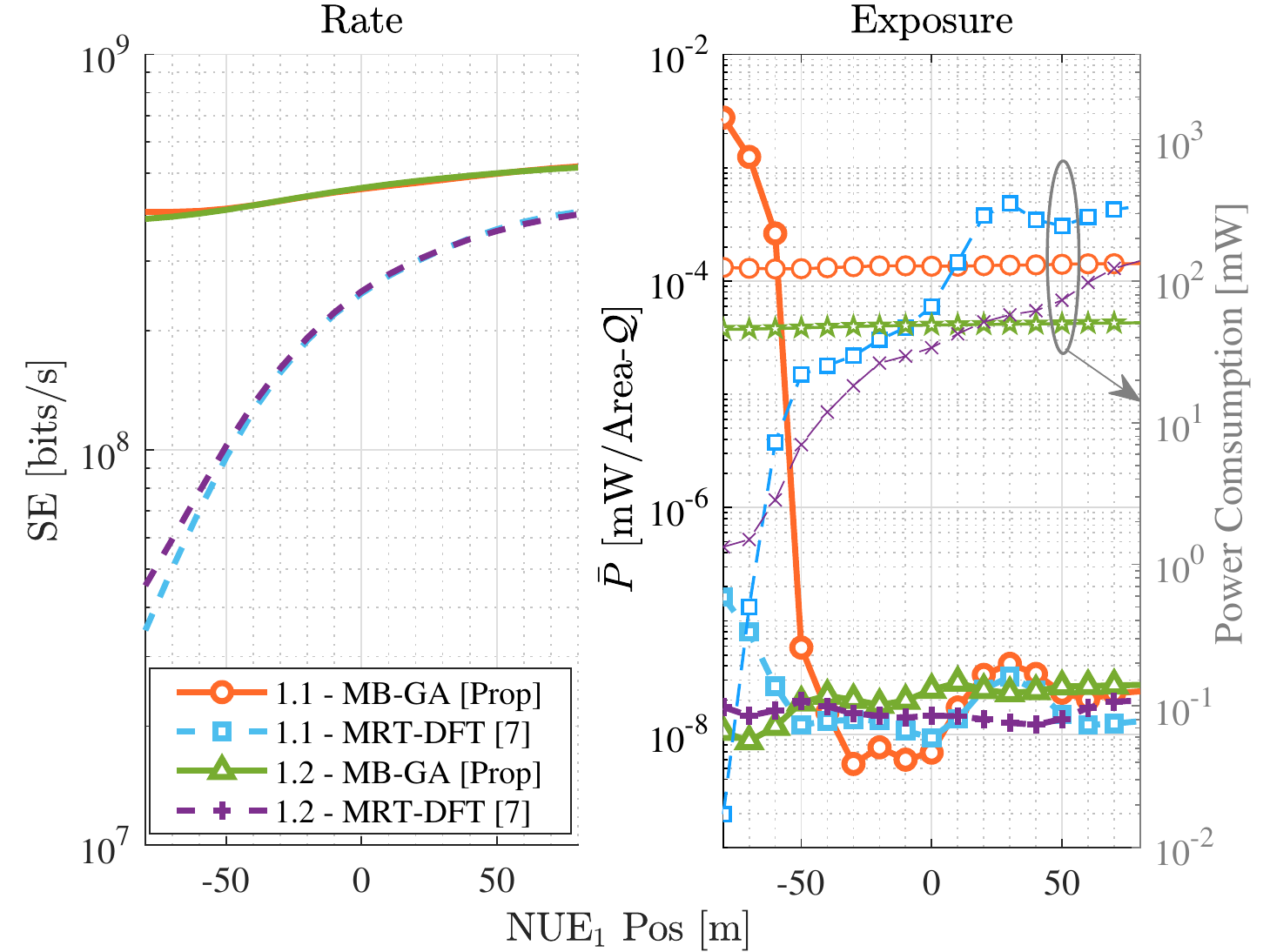}\vspace{-3mm}
\caption{Achieved rate \textit{(left)} and exposure \textit{(right)} performance under perfect \gls{csi} for both scenarios 2.1 and 2.2. The \gls{nue}$_1$ has the same positioning as in scenarios 1.1 and 1.2, and \gls{nue}$_2$ is positioned in \gls{bs}-\gls{ris} in steps of 5.494 meters.}
    \label{fig:2NUEPCSI}
\end{figure}

\noindent\textbf{Scenarios 2.1 and 2.2}: in these scenarios, the amount of power to be allocated is restricted in the both direct and reflected path since the \gls{emfe} constraint applies in both links. For the reference method, we apply the same rule of Algorithm \ref{ag:baseline} line 9 to the reflected path. Hence, the amount of power impinging in both \glspl{nue} is trackable. The results obtained for both scenarios are depicted in Fig. \ref{fig:2NUEPCSI}.
It shows a more significant performance gap in terms of rate. Both configurations of positioning{, considering BS, RIS, NUE, UE with equal or different heights,} achieved similar performances in terms of \gls{se}; however, the difference between the methods becomes clear. Compared to scenarios 1.1 and 1.2, the more truncated scenario highly limits the power, showing that the maximum achievable \gls{se} is half the value obtained earlier for the best case. The exposure of our method is higher, but the higher \gls{se} shows it is possible to manipulate the \gls{pa} vector further to reduce the exposure at the cost of some rate {degradation}. Also, both methods are able to compensate for the higher path-loss in \gls{3d} scenarios since the used power is very limited, up to only {0.3} W for both methods
. Due to the intrinsic characteristics of \gls{mmwave} \gls{mimo} systems, this ends up being a more {energy-}efficient transmission.

\subsection{Performance under Position Information}
\vspace{-1mm}
Here we evaluate the performance of the scenarios with limited information. Our goal is to check whether having only information about the position of the \gls{ue} and \glspl{nue}\footnote{We assume the \gls{bs} obtains this information via a sensing phase. The exact procedure is out of the scope of this paper and is left for future work.} impacts communication and \gls{emfe}. In Section \ref{sec:chestvsloc} we highlighted some drawbacks of using \gls{csi}, namely the overhead and self-exposure due to the \gls{chest} procedure necessary for both \gls{ue} and \glspl{nue}. 
To that end, we present here the results assuming the \gls{bs} has only the knowledge of \glspl{nue} and \glspl{ue} positions, \textit{i.e.} the array response of the users. The methods applied by the \gls{bs} are the same, but instead of using the channel vectors, the \gls{bs} computes a geometrical channel based on the positions of two arbitrary elements
 \cite{MARISA}. 
%
%
While a degradation in the performance is expected since the \gls{bs} is not aware of the multipath components that compose the channel coefficients, there is a benefit {in avoiding \gls{chest} procedure}.

\begin{figure}[htbp]
\centering
\includegraphics[width = .95\linewidth]{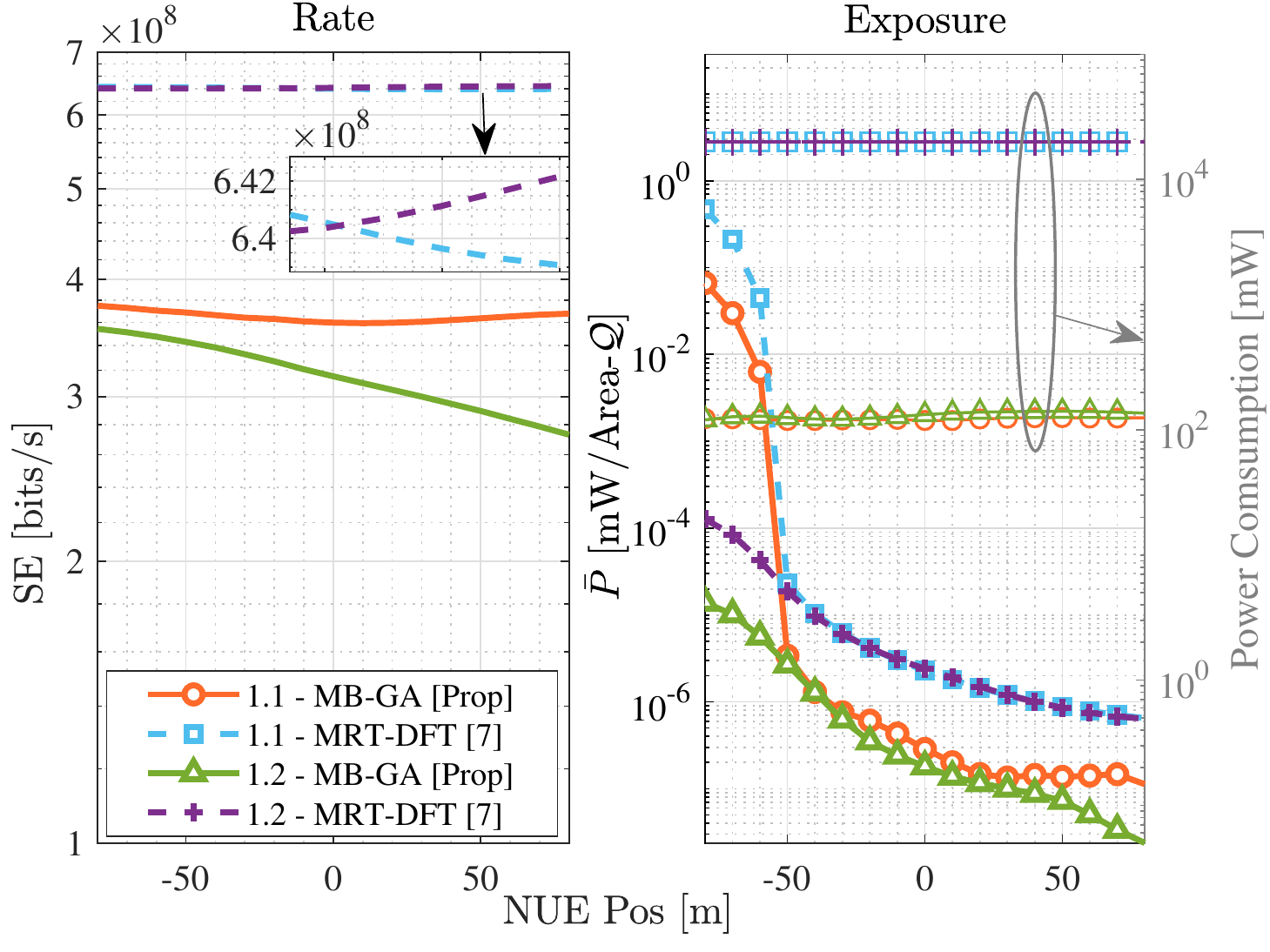}\vspace{-3mm}
\caption{{Achieved rate \textit{(left)} and exposure \textit{(right)} performance under localization for 1.1 and 1.2. The \gls{nue} is positioned in the origin of $y$-axis and move from $-80$ meters (\gls{bs} position) to $80$ meters (\gls{ue} position). 
}}
\label{fig:1NUELOC}
\end{figure}

\noindent \textbf{Scenarios 1.1 and 1.2}: We again establish the baseline for our solution with simpler configurations. The results for this setup are presented in Fig. \ref{fig:1NUELOC}. As opposed to the perfect \gls{csi} case, our approach yields a lower rate when compared to the reference method. However, this cost is offset {by a remarkable  reduction in the EMF}. Even when comparing the exposure levels with those achieved with perfect \gls{csi}, our method performs similarly and outperforms the reference approach. We would like to remark our approach prioritizes the reduction in \gls{emf}. 

\begin{figure}[t]
\centering
\includegraphics[width = .95\linewidth]{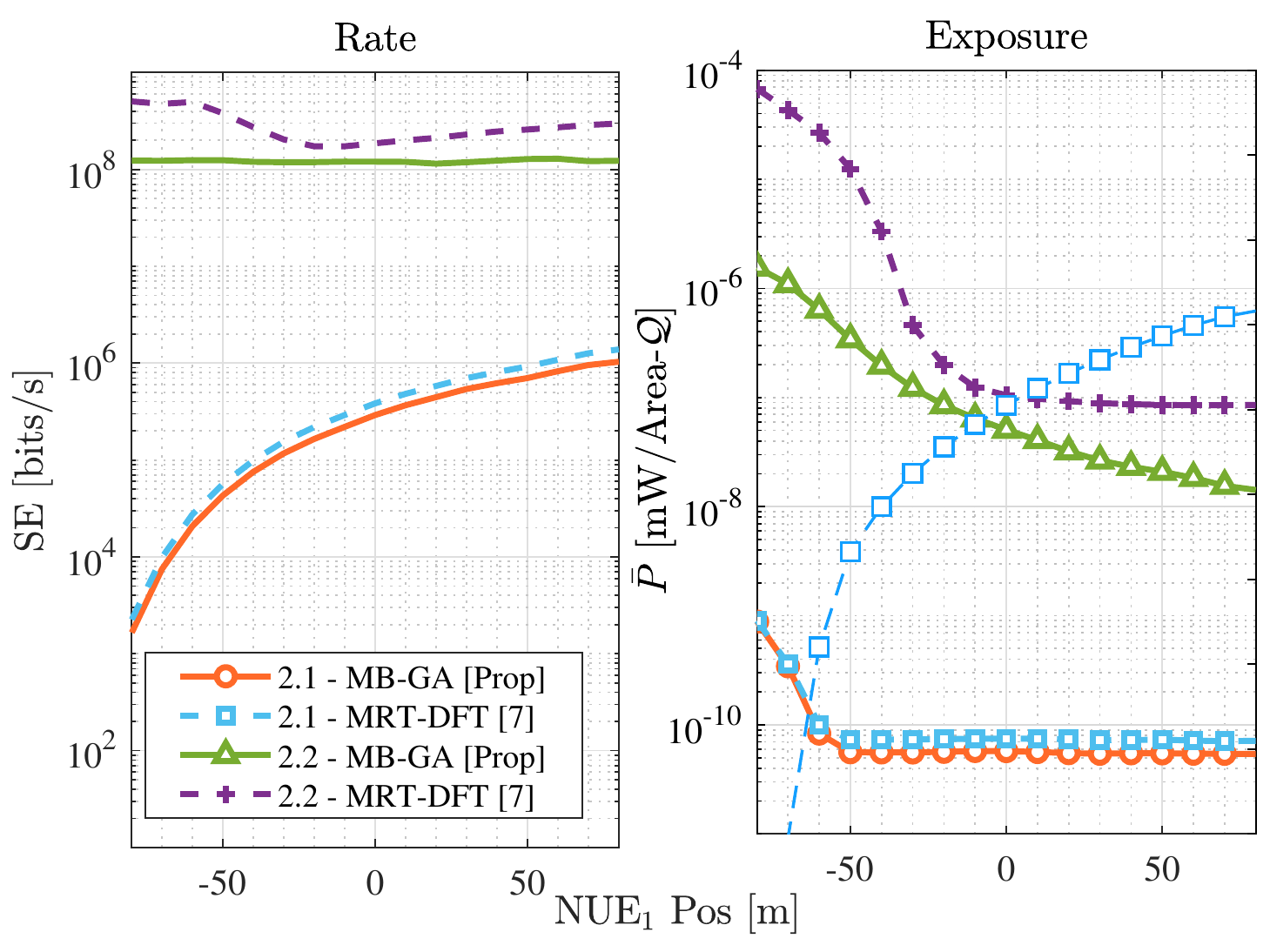}\vspace{-3mm}
\caption{{Achieved rate \textit{(left)} and exposure \textit{(right)} performance under localization for 2.1 and 2.2. The \gls{nue}$_1$ has the same positioning as in scenarios 1.1 and 1.2, and \gls{nue}$_2$ is positioned in \gls{bs}-\gls{ris} in steps of 5.494 meters.}} 
    \label{fig:2NUELOC}
\end{figure}

\vspace{2mm}
\noindent\textbf{Scenarios 2.1 and 2.2}:  
When compared to the perfect \gls{csi} {case, 
Fig. \ref{fig:2NUELOC} reveals} a similar rate performance between the two {EMF mitigation} methods, which can be attributed to the {refined power control attained} in \gls{mrt}-\gls{dft} as it accounts for the specific channel coefficients, {while the} MB-\gls{ga} is able to combine specific \gls{bf} vectors that sum non-coherently with small scale coefficients of the \glspl{nue}, enabling the use of more power to the transmission. 
In Scenario 2.1, the \gls{emfe} levels appear to be similar to those of referenced method. However, in the most realistic Scenario 2.2, 
{the proposed methods can attain substantial} exposure reductions of up to two orders of magnitude {w.r.t. \cite{HaoGuo2022}}. Importantly, our approach consistently delivers higher ratios of obtained rate to exposure {in $\frac{\rm bits/s}{\rm mW/m^2}$ } across all scenarios, irrespective of the perfect \gls{csi} or localization approach{es}. This underscores the efficacy of our solution even under the most challenging conditions.

\section{Conclusion and Future Works}

This work assessed the potential of multi-beam transmission with \gls{mmwave} directivity and \gls{ris} multi-path enhancement. Our results {reveal} that, thanks to the physical characteristics of next-generation technologies, it is possible to exploit multiple propagation paths to improve the quality of service and reduce \gls{emfe}. Specifically, assuming perfect \gls{csi}, the MB-\gls{ga} method overcomes another proposal in every metric at a complexity cost while assuming localization it has degraded \gls{se} performance but with lesser exposure. Future works include accounting for imperfect channel estimation and localization, as well as developing new algorithms to perform beam weighting. 

\vspace{-3mm}
\bibliographystyle{IEEEtran}
\bibliography{IEEEabrv,reference}
\end{document}